\documentclass[twocolumn,prl,superscriptaddress,showpacs,10pt,aps]{revtex4-1}

\newif\ifarxi   \arxifalse
\arxitrue  

\usepackage{amsmath}
\usepackage{amssymb}
\usepackage{graphicx}
\usepackage{color}

\usepackage[T1]{fontenc}
\usepackage[latin9]{inputenc}

\usepackage{amsthm}
\usepackage{bbm}
\usepackage{mathtools}

\def\epapps{\ifarxi appendix\else electronic supplement \cite{epaps}\fi}
\def\minisection#1{\ifarxi\section*{#1}\else\par\vskip20pt\noindent{\it #1.}\ \fi\noindent}
\def\minisubsection#1{\ifarxi\subsection*{#1}\else\par\vskip20pt\noindent{\it #1.}\ \fi\noindent}

\newtheorem{thm}{Theorem}

\newtheorem{lem}[thm]{Lemma}

\def\tr{\operatorname{tr}}
\def\idty{{\leavevmode\rm 1\mkern -5.4mu I}} 

\def\Cx{{\mathbb C}}
\def\Ir{{\mathbb Z}}\def\Nl{{\mathbb N}}

\def\norm #1{\Vert #1\Vert}
\def\Norm #1{\left\Vert #1\right\Vert}
\def\opNorm #1{\Norm{#1}_{\,{op}}} 
\def\mod{{\mathop{\rm mod}\nolimits}}

\def\ket #1{\vert#1\rangle}

\def\expect#1{\langle#1\rangle}
\def\tr{\mathop{\rm tr}\nolimits}
\def\abs#1{\vert#1\vert}

\def\Order{{\mathcal O}}

\let\rtu\eta  

\def\EF{\Phi} 
\def\nr{n} 
\def\dr{m} 
\def\Q{\hat{x}}  
\def\We{W_{\EF}}  
\def\Wz{W_{0}}  
\newcommand{\Wea}[1]{W_{#1}}  
\def\kk{i}   
\def\GR{\varphi} 

\begin{document}
\title{Propagation and spectral properties of quantum walks in electric fields}

\author{C. Cedzich}
\affiliation{Institut f\"ur Theoretische Physik, Leibniz Universit\"at Hannover, Appelstr. 2, 30167 Hannover, Germany}
\author{T. 
Ryb\'ar}
\affiliation{Institut f\"ur Theoretische Physik, Leibniz Universit\"at Hannover, Appelstr. 2, 30167 Hannover, Germany}
\author{A. H. Werner}
\affiliation{Institut f\"ur Theoretische Physik, Leibniz Universit\"at Hannover, Appelstr. 2, 30167 Hannover, Germany}
\author{A. Alberti}
\affiliation{Institut f\"ur Angewandte Physik, Universit\"at Bonn, Wegelerstr. 8, 53115 Bonn, Germany}
\author{M. Genske}
\affiliation{Institut f\"ur Angewandte Physik, Universit\"at Bonn, Wegelerstr. 8, 53115 Bonn, Germany}
\author{R. F. Werner}
\affiliation{Institut f\"ur Theoretische Physik, Leibniz Universit\"at Hannover, Appelstr. 2, 30167 Hannover, Germany}

\begin{abstract}
We study one-dimensional quantum walks in a homogenous electric field. The field is given by a phase which depends linearly on position and is applied after each step. The long time propagation properties of this system, such as revivals, ballistic expansion and Anderson localization,  depend very sensitively on the value of the electric field $\EF$, e.g., on whether $\EF/(2\pi)$ is rational or irrational. We relate these properties to the continued fraction expansion of the field. When the field is given only with finite accuracy, the beginning of the expansion allows analogous conclusions about the behavior on finite time scales.
\end{abstract}


\maketitle
\ifarxi\section*{Introduction}\fi

In this paper we analyze the surprisingly rich long time behavior of a simple quantum lattice system. Single neutral atoms in an optical lattice undergo a discrete time quantum walk by repeating the steps of first a unitary ``coin'' operation on the internal spin state, then a coherent spin-dependent transport by one lattice site and, finally, the application of a phase depending linearly on the lattice site. The last step, which corresponds to a homogeneous electric field, has recently been implemented by an acceleration of the lattice \cite{OurExp}. The crucial parameter is the phase difference per lattice site $\EF$. Depending on $\EF$,  quite different long time behaviors have been observed \cite{OurExp}, namely ballistic expansion, revivals, and Anderson localization.
The first  challenge is hence to identify those $\EF$ for which different infinite time behaviors and related spectral properties occur.
As we will see, this behavior is extremely discontinuous. It depends on whether $\EF/(2\pi)$ is rational or not, which can never be controlled or verified under finite error bars, and can therefore never be used to explain the differences seen in an experiment. However, if one does the experiment, and compares the behavior for $\EF/(2\pi)$ a simple fraction and, say, the (irrational) golden ratio one sees quite clearly already on small time scales the qualitative differences predicted for the infinite time behavior (cp. Fig.~\ref{fig:rat} and Fig.~\ref{fig:irrat} below). In this sense the rational vs.\ irrational distinction seems practically relevant after all. The second  challenge addressed in this paper is to resolve this apparent paradox, and to show how to extract qualitative information for large but not infinite times from a $\EF$ given with finite accuracy.

The most famous case in which the dependence of a spectrum on the rationality of a parameter was shown is the  Hofstadter butterfly \cite{Butterfliege}, which pertains to a two dimensional lattice system in a magnetic field. The crucial property is the rationality of the magnetic flux through one unit cell. This system is analyzed in terms of the one dimensional Mathieu equation, whose variants have recently been analyzed with great success in terms of certain dynamical systems \cite{Avila}. Interesting analogies to the present system are also provided by the quantum kicked rotator \cite{kicktherotor}. Exactly the same system has been studied mostly in the rational case, namely numerically in \cite{Banuls,Buerschaper} and by exact diagonalization on a finite ring \cite{Wojcik}.
In the irrational case Anderson localization \cite{AndersonReallyAnderson,Localization,AspectInguscio} is conjectured in \cite{Wojcik}. We agree that this is possibly the typical case assuming $\EF$ to be a uniformly distributed random number, but we also exhibit irrational cases with a quite different behavior. Most extremely, we show the possibility of ``hierarchical motion'', which is characterized by an infinite hierarchy of time scales in which ever more perfect revivals alternate with ever larger ballistic excursions.

An overview of our answers to the first challenge, the infinite time behavior, is given in Table~\ref{tab:sum}. Some of the terminology will be explained in detail in the respective sections.  The quasi-energy spectrum refers to the spectral properties of the unitary walk operator in the sense of, e.g., \cite[Ch.~VII.2]{ReedSimon1}.  The properties in the table depend on the exact value of $\EF$ or, equivalently on its entire continued fraction representation.
\begin{table}[h]
\begin{tabular}{c||c|c|c}
       &rational&\ almost rational&very irrational\cr\hline
  cont. fract.&terminates& $c_i\to\infty$ & $c_i$ bounded \cr
  expansion &           &   rapidly \cr\hline
  propagation& ballistic&hierarchical&localized\cr
         &w. revivals    & &almost periodic \cr\hline
  quasi-energy& absolutely& singular& dense set of\cr
  spectrum&continuous&continuous&eigenvalues\cr\hline
   status & proved   &proved& num. evidence\cr
\end{tabular}
\caption{Overview: The connection between the properties of $\EF$, its continued fraction coefficients $c_i$, propagation, and the spectrum of the unitary evolution operator. Detailed explanations are found in the subsections for the respective columns.}
\label{tab:sum}
\end{table}
For an experimentally given field with finite accuracy, only an initial segment of this expansion can be reliably determined. Depending on how one continues the expansion, any of the three classes is compatible with such data. However, the behavior on a finite time scale can be reliable predicted from the data. As in the infinite case the important input is the size of the continued fraction terms $c_i$, which, in particular, determine the quality of revivals. These relationships constitute the answer we can give to the second challenge.

\minisection{Definition of the system} For definiteness and to fix notations let us explicitly define the class of systems under investigation. We consider particles on the 1D lattice $\Ir$ with a two dimensional space of internal degrees of freedom. Basis vectors are thus $\ket{x,\alpha}$ with $x\in\Ir$ and $\alpha=\pm1$. The state dependent shift is defined by $S\ket{x,\alpha}=\ket{x+\alpha,\alpha}$. We consider, moreover, a coin operation
$C\ket{x,\alpha}=\sum_\beta C_{\alpha\beta}\ket{x,\beta}$ with a fixed matrix $C\in {\rm SU}(2)$, i.e.,
\begin{equation}\label{coin}
    C=\left(\begin{array}{cc}a&b\\-\overline{b}&\overline{a}\end{array}\right)
\end{equation}
with $\abs a^2+\abs b^2=1$. A standard case is the Hadamard walk with $a=b=1/\sqrt2$. The third element will be a phase shift depending linearly on $x$, given by the operator $\exp(i\Q\EF)\ket{x,\alpha}=\exp(ix\EF)\ket{x,\alpha}$. The constant $\EF$ is referred to as the electric field. The overall time step is then given by
\begin{equation}\label{WE}
    \We=e^{i\Q\EF}CS\ .
\end{equation}

\minisection{Rational approximations}
We will analyze the behavior of $\We$ and its iterations by comparison with similar walks $\Wea{{\EF}'}$, where $\EF'=2\pi \nr/\dr$ is a rational approximation to $\EF$.
The basic tool is an estimate for the difference between two states evolved from the same initial state by walks with slightly different fields $\EF$ and $\EF'$.  As initial state  we choose a unit vector $\psi$ which is non-zero only for lattice sites $x$ with $\abs{x}< L$.  Then for one step $\norm{\We\psi-\Wea{{\EF}'}\psi}\leq\max_{\abs x\leq L}\abs{\exp(ix\EF)-\exp{ix\EF'}}\leq L\abs{\EF-\EF'}$. After $t$ steps, adding the deviations, and observing that the localization region of $\psi$ increases by $1$ in every step we get:
\begin{equation}\label{tstepsEEprime}
    \norm{\We^t\psi-\Wea{\EF'}^t\psi}\leq \frac t2(t+2L-1)\ \abs{\EF-\EF'}\ .
\end{equation}
So, for example, if $\EF'$ approximates $\EF$ to within $\varepsilon=10^{-4}$, and the particle starts at the origin ($L=1$), we can use $\Wea{{\EF}'}^t\psi$ to predict the behavior of $\We^t\psi$ for up to $\varepsilon^{-1/2}\approx100$ steps. The square root makes decimal approximations (or approximations in any other digital base) quite useless. Instead, one can look for denominators $\dr$ such the approximation error is not of order $\dr^{-1}$, as can always be trivially achieved, but of order $\dr^{-2}$ or better.

The systematic way to generate such approximations is the so-called continued fraction expansion \cite{Hardy}.
For a number $x=x_0\geq0$ it is computed by taking the integer part $c_0=\lfloor x_0\rfloor$, setting $x_1=1/(x-c_0)$, and repeating this process for $x_1$ to get $c_1$, etc. The number $x$ is then uniquely characterized by the sequence $(c_0,c_1,\ldots)$. One gets an explicit sequence of approximating rationals ${\nr}_{\kk}/{\dr}_{\kk}$, where the denominators ${\dr}_{\kk}$ and numerators ${\nr}_{\kk}$ both satisfy the recursion $r_{{\kk}}=c_{\kk}r_{{\kk}-1}+r_{{\kk}-2}$ with initial values ${\nr}_0=c_0$, ${\nr}_{-1}=1$, ${\dr}_0=1$, ${\dr}_{-1}=0$. This is equivalent to an iteration in which at step $i$ the denominator $c_i$ is replaced by $c_i+1/c_{i+1}$, leading to the typographical nightmare from which the continued fractions derive their name. The desired approximation of $x$ of order $1/{\dr}_{\kk}^2$ is stated as
\begin{equation}\label{cfapprox}
    \left|x-\frac{{\nr}_{\kk}}{{\dr}_{\kk}}\right| < \frac1{c_{{\kk}+1}{\dr}_{\kk}^2}.
\end{equation}
Therefore, when $\EF'=2\pi\,\nr_{\kk}/{\dr}_{\kk}$ is a continued fraction approximation of $\EF$, the error in \eqref{tstepsEEprime} is of order $(t/\dr_i)^2$, so the $i^{\text{th}}$ approximation is valid roughly on the time scale of the denominator $\dr_i$, and correspondingly longer if the next term $c_{i+1}$ is large.

Intuitively, equation \eqref{cfapprox} says that large terms $c_i$ correspond to especially good approximations. For example, the continued fraction expansion of $\pi$ begins with $(3,7,15,...)$ and the second approximant $\pi\approx22/7$ has an error $<1/(15*49)\approx10^{-3}$.
In this sense the ``most irrational number'' is the golden ratio $\GR=(1+\sqrt5)/2$ with expansion $(1,1,1,\ldots)$.

\minisection{The rational case: Revivals and ballistic expansion}
Let us begin with the rational case, in which $\EF=2\pi\,\nr/\dr$, where it is understood that $\nr,\dr$ do not have common factors. The phase factor $\rtu=\exp(i\EF)$ is then a primitive $\dr^{\rm th}$ root of unity, i.e., $\rtu^{\dr}=1$, but this holds for no smaller power. There are two ways to reduce the analysis to the case of standard, translationally invariant walks \cite{Grimmett,TRcoin,SpaceTimeCoinFlux}. One is to use the fact the field phase factor is periodic with period $\dr$, and hence always group $\dr$ sites together into a ``supercell'' of ``coin dimension'' $2^\dr$. This spatial regrouping is carried out in \cite{OurExp}, giving a band structure with a $2^\dr$-branched dispersion relation. Here we will use instead a temporal regrouping, i.e., consider the operator $\We^{\dr}$, which likewise commutes with translations. This is a walk with two coin states, but allowing for steps over up to $\dr$ sites.
The key fact about this walk is that, even for moderately large $\dr$, it hardly moves, in the sense that the walk operator is close to the identity operator (up to a phase). Phrased as a statement about the ungrouped walk $\We$, we get a revival of the initial state after $\dr$ steps. The precise statement is the following {\it revival theorem}:
\begin{alignat}{2}\label{reviveOdd}
    \opNorm{\,\We^{2\dr}+\idty\,}&= 2\abs{a}^\dr& \qquad&\dr\text{ odd} \\
    \opNorm{\,\We^{\dr}+(-1)^{\dr/2}\idty\,}&= 2\abs{a}^{\dr/2}& &\dr\text{ even},
    \label{reviveEven}
\end{alignat}
Hence for odd $\dr$ {\it any} initial state is reproduced after $2\dr$ steps with accuracy (trace norm distance) $4\abs{a}^{\dr}$, which is exponentially small in $\dr$, since $\abs{a}<1$, except in trivial cases. For the Hadamard walk we get an error $2^{-\dr/2}$. The different behavior  of the even and odd cases is to be expected for coined walks, since the amplitude for no jump or a jump by an even distance is zero. Hence the revival of a state localized at a point is only possible after an even number of steps.

The proof of the revival theorem will be given in the \epapps, and could also be based on the eigenvalues given, without proof,  in \cite{Wojcik}. It yields also the dispersion relations for $\We^\dr$, concretely $\cos\omega_\pm(k)=$
\begin{equation}\label{dispregroup}
  =\!\left\lbrace\!\! \begin{array}{cl}a^{\dr}\cos(\dr k)& \dr\mbox{\ odd}\\
                               -a^{\dr}\cos(\dr k)+(-1)^{{\dr}/2+1}(1-\abs a^\dr) &  \dr\mbox{ even}
                              \end{array}\right.
\end{equation}
The upshot is that the spectrum is absolutely continuous, and transport is ultimately ballistic. However, because of the revivals it may take a very long time for the ballistic regime to be reached. This is shown in Fig.~\ref{fig:rat} for walks starting at the origin.
In each case we show two aspects as a function of time, the root mean square deviation of position $\sigma(t)=\langle x^2\rangle^{1/2}$ and the probability $p(t)$ of return to the origin. Perfect revivals are identified by the conditions $\sigma(t)=0$, resp.\ $p(t)=1$. The two fields chosen in the figure are $\EF_1=2\pi/5$, for which \eqref{reviveOdd} predicts a fairly weak revival at $t=10$ ($p(10)\geq.64$), which rapidly goes over into ballistic expansion. The envelope of $p(t)$ is well approximated by a Bessel function \cite{Wojcik}.
The denominator $\EF_2=2\pi\times 51/256$ for the second field is much larger, so the revival predicted by \eqref{reviveEven} is exponentially sharper: $p(256)\geq1-10^{-19}$. The full evolution up to $t=256$ therefore repeats roughly $10^{19}$ times, i.e., until the revival errors  accumulate sufficiently, to make way for ballistic expansion. This would be true for all fields of the form $\EF=2\pi\nr/256$, independently of the numerator $\nr$. However, the evolution up to $t=256$ does depend on the numerator \cite{Banuls}. We can get some information about it from a ``rational approximation of the rational'' $51/256$: Its continued fraction sequence is $(c_0,c_1,c_2)=(0,5,51)$ and $1/5=\EF_1/(2\pi)$. Therefore, we know that the initial segments of these evolutions coincide. More precisely, from \eqref{tstepsEEprime} we get $\norm{\Wea{\EF_2}^t\psi-\Wea{{\EF_1}}^t\psi}\leq t(t+1)/t_0^2$ with $t_0\approx 20$. The comparison for the displayed quantities is even more favorable, as shown in the insert of Fig.~\ref{fig:rat}.

\begin{figure}[ht]
  \includegraphics[width=\columnwidth]{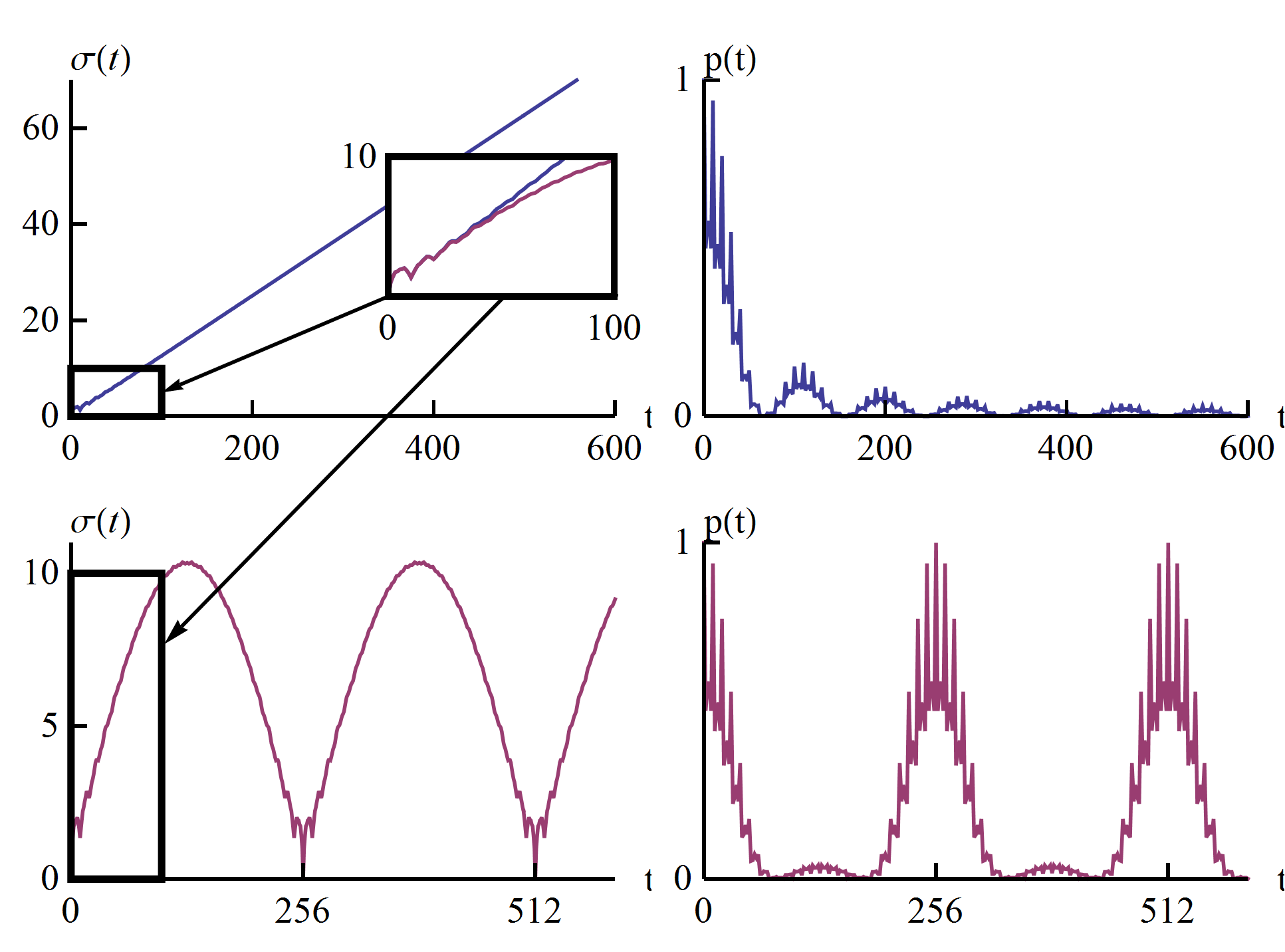}
  \caption{Revivals for two walks with rational fields. Top $\EF_1=2\pi/5$ and bottom $\EF_2=2\pi\times 51/256$. In both cases the right panel shows the root mean square of the position, and the left panel the probability $p(t)$ to be back at the origin at time step $t$. Since $\EF_1-\EF_2\approx5*10^{-3}$ the initial segments ($t<100$) of the top and bottom graphs almost coincide (see black frames and insert). The revivals predicted by the revival theorem are at $t=10$ and $t=256$, respectively. }
  \label{fig:rat}
\end{figure}

\minisection{The almost rational case}
The same idea, i.e., combining \eqref{tstepsEEprime},\eqref{cfapprox}, can be used with the revival theorem \eqref{reviveOdd} resp.\ \eqref{reviveEven} to get revival estimates for arbitrary irrational $\EF$:
\begin{equation}\begin{aligned}\label{reviveirr}
    \Norm{\,\We^{2{\dr}_{\kk}}\psi+\psi\,}&\leq \frac{4\pi}{c_{\kk+1}}+\Order\left(\frac L{{\dr}_{\kk}}\right) \quad\dr_{\kk} \text{ odd}   \\
    \Norm{\,\We^{\dr_{\kk}}\psi+(-1)^{\dr_{\kk}/2}\psi\,}&\leq \frac{\pi}{c_{\kk+1}}+\Order\left(\frac L{\dr_{\kk}}\right)  \quad\dr_{\kk}\text{ even}.
\end{aligned}\end{equation}
That is, if $c_{\kk_\alpha+1}\to\infty$ along some subsequence $\kk_\alpha$ we get an infinite sequence of sharper and sharper revivals at the corresponding times $\dr_{\kk_\alpha}$ or $2\dr_{\kk_\alpha}$. From each of these revivals we also get a more precise repetition of the entire history up to that point. This shows that the walk operator now has no absolutely continuous spectrum, since for initial vectors in the absolutely continuous subspace the expectation for any finite dimensional projection goes to zero by the Riemann-Lebesgue lemma.

When the $c_i$ grows sufficiently fast, we get hierarchical motion on a hierarchy of time scales $\dr_\kk$, each of which is associated with one approximation $\nr_\kk/\dr_\kk$. Sharp revivals, up to errors $\varepsilon_\kk$ with $\varepsilon_\kk{\to}0$, alternate with large excursions, whose length is given by a sequence of increasing intervals $I_i$.
The walk will have the property that for any state initially supported by $I_\kk$ there is a time when the probability to be back in $I_\kk$ is less than $\varepsilon_\kk$. The sequences $\varepsilon_\kk$ and $I_\kk$ can be prescribed arbitrarily.

We now show how a walk with these properties can be constructed by judiciously fixing the sequence $(c_0,c_1,\ldots)$ step by step. Suppose the sequence is constructed up to $c_\kk$. Then a revival is predicted by \eqref{reviveirr} at $\dr_\kk$ or $2\dr_\kk$. Clearly, by choosing $c_{\kk+1}$ sufficiently large we can make the error in \eqref{reviveirr} less than $\varepsilon_\kk$.  Now the rational walk with $\EF_\kk=2\pi\nr_\kk/\dr_\kk$ will eventually turn ballistic. That is, there is an ``excursion time'' at which any initial state supported on $I_\kk$ will be back in $I_\kk$ with probability less than $\varepsilon_\kk/2$. Choosing $c_{i+1}$ even larger if necessary we can ensure by \eqref{tstepsEEprime} and \eqref{cfapprox} that the approximation will hold up to this excursion time with error less than $\varepsilon_\kk/2$. Hence the constructed walk will have the desired excursion property.

We remark that in this construction the $c_\kk$ will have to grow very fast indeed, since after a sharp revival the ballistic stage is only reached after many repetitions. The fields $\EF$ arising from our construction are therefore not typical (the set has Lebesgue measure zero).   Hierarchical motion implies that the spectrum of $\We$ is purely singular continuous. In fact, we have already excluded absolutely continuous spectrum as inconsistent with infinitely many revivals. On the other hand there can be no eigenvalue (point spectrum). Indeed an eigenvector would be essentially (up to some $\varepsilon$) supported by some $I_\kk$. Since an eigenvector remains constant in time under the walk this contradicts the excursion property.

\minisection{The very irrational case}
Numbers with bounded continued fraction sequence $c_{\kk}$ do not have exceptionally good rational approximations. For example, for the golden ratio $\GR$ with $c_{\kk}=1$ for all ${\kk}$, the bound \eqref{reviveirr} makes no revival predictions at all. Fig.~\ref{fig:irrat} shows the behavior of root mean square position and return probability for $\EF=2\pi\GR$. It shows many peaks, not just at the continued fraction denominators, and no tendency for transport or decaying return probability. This is precisely the behavior one would expect from a quantum system with pure point spectrum. Indeed a numerical study readily confirms this guess.

\begin{figure}[ht]
  \includegraphics[width=8.2cm]{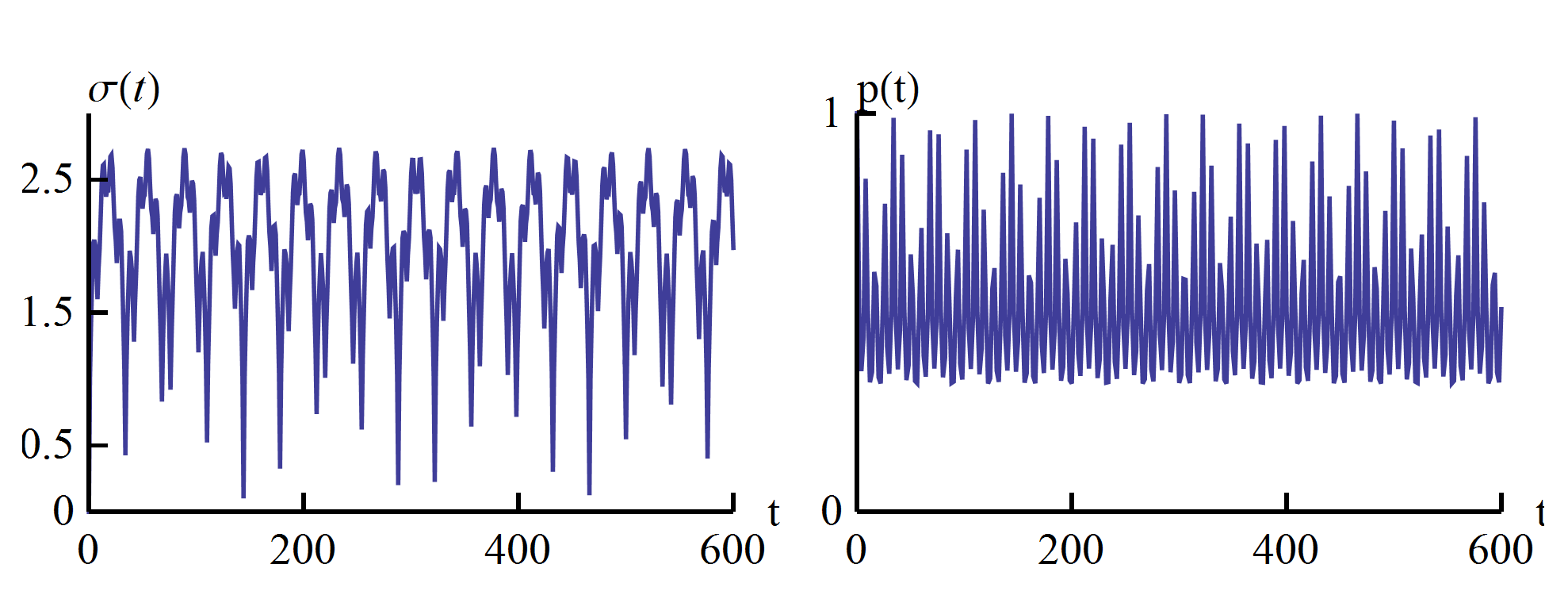}
  \caption{Root mean square of position and return probability for the golden ratio field $\EF=\pi(\sqrt5-1)$. There are many returns, and the initial state has a substantial overlap with a bound state, so the return probability is bounded away from zero. Vertical lines: Times appearing in the revival estimate, even though the estimate is trivial. }
  \label{fig:irrat}
\end{figure}

\begin{figure}[h]
  \includegraphics[width=8.2cm]{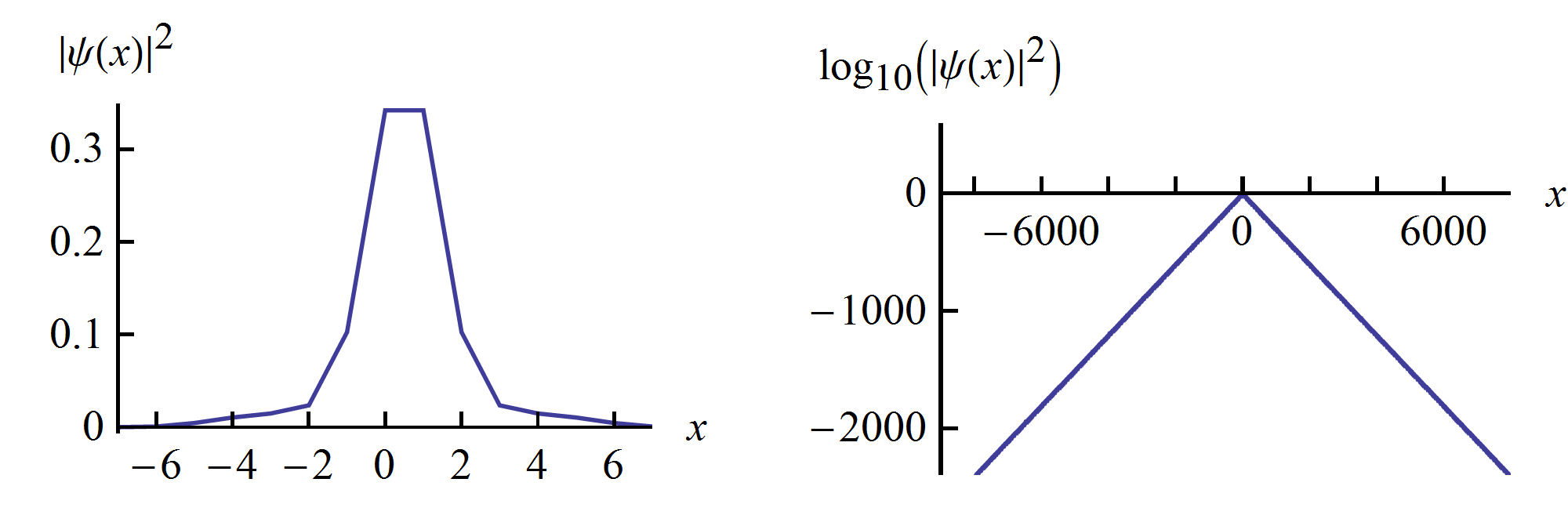}
  \caption{Position probability distribution for the eigenfunction of the walk with $\EF=2\pi\GR$. Left: Unscaled plot. Right: Logarithmic plot over the range $[-N,N]$ for which the procedure described in the text was applied.
  Floating point accuracy for this computation was set to 8000 decimal digits.
  The slope of the right graph is the inverse Anderson localization length $\approx.301$. }
  \label{fig:efunc}
\end{figure}

More precisely, we take the system with periodic boundary conditions, diagonalize it, and pick the eigenfunction with the smallest  $\expect{x^2}$. This eigenfunction decays exponentially (as long as one does not approach the boundary), and  converges as the ring size is increased. The same function is obtained by iterating the transfer matrices \cite{Localization} for the solution of the eigenvalue equation $\We\psi=\exp(i\omega)\psi$. The quasi-energy $\omega$ is chosen as $\omega=\EF/2$ as this ensures the symmetry of the eigenfunctions around the origin. Therefore, if we start with an arbitrary vector at some very negative site $x=-N$, the iteration will naturally pick the expanding branch, but on passing $x=0$ will decrease again to reach exactly the starting value at $x=+N$. When $N$ is large, this computation has to be carried out with high numerical precision (see Fig.~\ref{fig:efunc}). Setting the function to be zero for $\abs x>N$ yields an approximate eigenfunction, which solves the eigenvalue equation with a precision (e.g., $2400$ digits for the case shown in Fig.\ref{fig:efunc}), which in ordinary numerical work would be considered extraordinary high.

Knowing one exact eigenfunction is knowing a complete set, since we can generate the required number per site by shifts and staggered sign changes $\psi(x)\mapsto(-1)^x\psi(x)$. We emphasize that in spite of the high precision the numerical evaluation does not prove the eigenvalue equation, since for larger $N$ the behavior could be different, and indeed will be different for rationals close enough to $\EF$. Hence we have shown Anderson localization ``only for all practical purposes'', meaning for runs with less than $\sim10^{2000}$ steps, which for nanosecond steps will be incomparably larger than the age of the universe.

It is paradoxical that proving localization for a concretely specified irrational is very hard, whereas it is possibly much easier to show that
localization holds for all $\Phi$ outside a set of measure zero. There are some indications for the latter statement to be true. Indeed one can repeat the above eigenvector computation for random $\Phi$ and finds that the inverse localization lengths are always around $.301$ with a variance that decreases when $N$ and the numerical accuracy are increased. In contrast, the small scale behavior (left panel in Fig.~\ref{fig:efunc}) of the eigenfunction does depend visibly on $\Phi$. We note that this conjecture does not contradict the markedly different behavior for rational and almost rational $\EF$, since these sets are of measure zero.

\minisection{Summary and outlook}
We have established three typical kinds of long time behavior of electric walks as a function of the  field $\EF$. This dependence is extremely discontinuous in the sense that an arbitrarily small change in $\EF$ may change the type of propagation behavior in the infinite time limit. Nevertheless, on a given time scale one can make precise predictions in terms of an initial segment of the continued fraction approximation of $\EF$. Anderson localization holds for very irrational numbers, and we conjecture that it also holds for random numbers $\EF$ with probability one with a non-random localization length.

\ifarxi\section*{Appendix}The revival theorem is stated in equations  \eqref{reviveOdd} and \eqref{reviveEven} in the body of the paper. We repeat it here for reference:

\begin{thm}
Let $\Phi=2\pi\frac{n}{m}$ such that $n$ and $m$ are relatively prime integers. Then the electric walk operator $\We$ satisfies the revival relations
\begin{alignat}{2}\label{reviveOdd_app}
    \opNorm{\,\We^{2\dr}+\idty\,}&= 2\abs{a}^\dr& \qquad&\dr\text{ odd} \\
    \opNorm{\,\We^{\dr}+(-1)^{\dr/2}\idty\,}&= 2\abs{a}^{\dr/2}& &\dr\text{ even}\; .
    \label{reviveEven_app}
\end{alignat}
\end{thm}

In order to prove this theorem, we analyze the operator $\We$ in momentum space. That is, we consider wave functions on the Brillouin zone with values in the coin space $\Cx^2$. Operators commuting with translations are then precisely those which act by multiplication with a matrix depending on the quasi-momentum $k$. In particular, the coined walk in zero field acts by
$\Wz(k)\hat\psi(k)$,
\begin{equation}\label{Wp}
    \Wz(k)=S(k)C=\left(\begin{array}{cc}ae^{ik}&be^{ik}\\-\overline{b}e^{-ik}&\overline{a}e^{-ik}\end{array}\right)
    \ ,
\end{equation}
where $S(k)=\exp{ik\sigma_3}$.
The operators $e^{i\Q\EF}$ act in momentum space as a shifts of the quasi momentum, i.e., $(e^{i\Q\EF}\hat\psi)(k)=\hat\psi(k+\EF)$ where the addition is understood $\mod\,2\pi$, as usual for quasi momentum, so in particular $((e^{i\Q\EF})^m\hat\psi)(k)=\hat\psi(k)$. Applying $\We$ therefore gives
$(\We\hat\psi)(k)=\Wz(k+\EF)\hat\psi(k+\EF)$. With $\dr$ applications the momentum is shifted by $\dr\EF$, which by assumption is a multiple of $2\pi$, and can hence be left out. This precisely reflects the anticipated translation invariance of $\We^{\dr}$. Hence, this operator is represented by the $k$-dependent $2\times2$-matrix
\begin{eqnarray}\label{WEnp}
    \We^{\dr}(k)&=&\Wz(k+\EF)\cdot \Wz(k+2\EF)\cdots  W(k+\dr\EF)   \nonumber\\
            &=& S(\EF)\Wz(k)S(\EF)^2\cdots S(\EF)^{\dr}\Wz(k)\ .
\end{eqnarray}
By definition, the dispersion relations associated with such a walk are the functions $\omega_\pm(k)$ such that the eigenvalues of $\We^{\dr}(k)$ are $\exp(i\omega_\pm(k))$.  Since we are dealing with $2\times 2$ matrices and $\det(\We^{\dr}(k))=1$ the dispersion relations depend only on the trace of $\We^{\dr}(k)$, more precisely, we have
\begin{equation}\label{protodispersion}
  2\cos\omega_\pm(k)=\tr(\We^{\dr}(k))\;.
\end{equation}
Evaluating this trace could be expected to be a hard task involving a sum over $2^{\dr}$ oscillating terms. Surprisingly, however, there is a simple formula for this type of expression. We state it slightly more generally than needed. The proof of the Theorem will be completed after the proof of this Lemma.

\begin{lem}\abovedisplayskip=3pt\belowdisplayskip=3pt
Let $m\in\Nl$, and $\eta$ be a primitive $m^{th}$ root of unity. Consider the matrices
\begin{equation}\label{CandR}
   C=\left(\begin{array}{cc}a&b\\c&d\end{array}\right) \quad\text{and}\quad
   R=\left(\begin{array}{cc}\eta&0\\0&\eta^{-1}\end{array}\right),
\end{equation}
and set
\begin{equation}\label{tracexpr}
    \tau_m(C) = \tr(CR^0CR^1\cdots CR^{m-1})\;.
\end{equation}
Then for odd $m$:
\begin{equation}\label{trodd}\nonumber
 \tau_m(C)=a^m+d^m\;,
\end{equation}
and for even $m$:
\begin{equation}\label{treven}\nonumber
 \tau_m(C)=-\bigl(a^{m}+d^{m}\bigr)
          -2\bigl((-ad)^{m/2}-\det(C)^{m/2}\bigr).
\end{equation}
\end{lem}

\minisubsection{Proof of the Lemma}
For any $n$ (not necessarily $n=m$), consider the expression  $\tau_{n}(C)$ written out as the sum,
\begin{equation}
  \tau_{n}(C)=\sum_{\mathclap{i_1\dots i_n}}C_{i_1i_2}R^0_{i_2i_2}C_{i_2i_3}\dots C_{i_n i_1}R^{n-1}_{i_1i_1}.
\end{equation}
Then with each term we can associate a closed path $i_1\to i_2\to\cdots\to i_n\to i_1$ of length $n$ on the graph
\begin{center}
\includegraphics[width=\columnwidth]{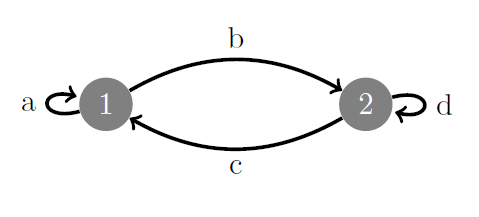}
\end{center}
This means that every summand contains the same number of $b$ and $c$ factors, so $\tau_n(C)$ depends only on the product $bc$.
Hence we can write it as a polynomial in the variables $a, d$ and $bc$;
\begin{equation}\label{taunSum}
   \tau_n(C) =  \sum_{rs}\gamma_{rs}\, a^r d^s\, (bc)^{(n-r-s)/2} \; ,
\end{equation}
with suitable coefficients $\gamma_{rs}$, where the degree of $(bc)$ follows from the homogeneity of degree $n$ of $\tau_n(C)$ in all four variables $(a,b,c,d)$.

Now assume that $m$ is odd, set $n=m$ in \eqref{taunSum}, and substitute $C\mapsto RC$.
This does not change the trace because $R^{m}=\idty=R^0$ and we can cyclically shift the factors $CR^k$ under the trace. On the other hand, this substitution replaces $a\mapsto\eta a$ and $d\mapsto \eta^{-1}d$. Thus the $r,s$-term in \eqref{taunSum}  gets multiplied by $\eta^{(r-s)}$. Since $\tau_m(C)$ is unchanged, the uniqueness of the polynomial coefficients tells us that the only contributing monomials can be those with $r\equiv s\ \mod\,m$. Since $0\leq r,s\leq m$ it follows that either $r=m,s=0$, $r=0,s=m$, or $r=s$. However, the last possibility is ruled out, because the overall degree of the trace as a polynomial in $(a,b,c,d)$ is $m$, and hence odd. Therefore the only non-zero coefficients are $\gamma_{0m}$ and $\gamma_{0m}$. Clearly, these coefficients are just the product of the respective powers of $R$, namely $\eta^N$ with $N=\sum_{k=0}^{m-1}k=m(m-1)/2$. Since $(m-1)/2$ is an integer, we have $\eta^N=1$, which entails the expression given in the Lemma.

The argument in the even case is similar. Let us set $m=2{m}'$. Our first aim is to compute $\tau_{m'}(C)$.
We introduce the operators
\begin{align}\label{eq:x_ell}
  X_\ell=CR^\ell\cdots CR^{\ell+{m}'-1}\; ,
\end{align}
i.e. products involving a contiguous block of only half the terms in the trace. By definition,
\begin{equation}\label{tauCtrX}
  \tr(X_\ell)=\tau_{m'}(CR^\ell)\;.
\end{equation}
From $\eta^m=1=(\eta^{m'})^2$ we get $\eta^{m'}=\pm1$. But since $\eta$ is a primitive $m^{th}$ root,  $\eta^{{m}'}=-1$, and $R^{{m}'}=-\idty$.
Inserting this relation into the trace of $X_1$ we find the relation
\begin{align}
  \tr(X_1) &=\tr(CRCR^2\cdots CR^{m'-1}CR^{m'}) \nonumber\\
           &=\tr(CR^{m'}CRCR^2\cdots CR^{m'-1})  \nonumber\\
           &=-\tr(X_0)\; .
\end{align}
\begin{equation}\label{tau0even}
  \tau_{m'}(CR)+\tau_{m'}(CR^2)=0\;.
\end{equation}
With \eqref{taunSum}, we see that the polynomial on the left hand side can only vanish if, for all $r,s$,
\begin{equation}\label{gam0even}
  \gamma_{rs}\bigl(\eta^{r-s}+1\bigr)=0.
\end{equation}
Since $-m'\leq r-s\leq m'$, and the only integer powers $t\in[-m',m']$ with $\eta^t=-1$ are $t=\pm m'$, we conclude that, as in the odd case, $\gamma_{rs}\neq0$ only for
$(r,s)=(m',0)$ or $(r,s)=(0,m')$. Also as in the odd case $\gamma_{m'0}=\eta^N$ with $N=\sum_{k=0}^{m'-1}k=m'(m'-1)/2$. We know that $\eta^{m'}=-1$ so we can write
$\gamma_{m'0}=j^{m'-1}$ and $\gamma_{0m'}=(-j)^{m'-1}$ with $j=\pm i$. Which sign is correct depends on the root $\eta$: We can write the primitive root as $\eta=\exp(2\pi i n/m)=\exp(i\pi n/m')$ for some odd $n$ coprime with $m'$.
Then $\eta^N=\exp\bigl((i\pi n/2)(m'-1)\bigr)$, i.e., $j=i^n$. To summarize:
\begin{equation}\label{taumprime}
  \tau_{m'}(C)= j^{m'-1}a^{m'}+(-j)^{m'-1}d^{m'} \;.
\end{equation}
Later on we will need only the square of this trace, so the ambiguity about the sign of $j$ drops out:
\begin{equation}\label{taumprime2}
  \tau_{m'}(C)^2= (-1)^{m'-1}\bigl(a^{m}+d^{m}\bigr)+2a^{m'}d^{m'} \;.
\end{equation}

Now the full trace we want to evaluate is $\tau_m(C)=\tr(X_0X_{m'})$ and since $X_{{m}'}=(-1)^{{m}'}X_0$ this can be done by the identity $\tr(X)^2-\tr(X^2)=2\det(X)$ for $2\times2$-matrices:
\begin{align}
  \tr(X_0 X_{m'})&= (-1)^{{m}'}\tr(X_0^2)      \nonumber\\
  &=(-1)^{{m}'}\Bigl((\tr X_0)^2-2\det(X_0) \Bigr) \nonumber\\
  &=(-1)^{{m}'}\Bigl(\tau_{m'}(C)^2-2\det(C)^{m'} \Bigr) \nonumber\\
  &=-\bigl(a^{m}+d^{m}\bigr)-2\bigl((-ad)^{m'}-\det(C)^{m'}\bigr)\nonumber
\end{align}
which finishes the proof of the Lemma.

\minisubsection{The dispersion relation}
Coming back to the walk in a rational electric field we can evaluate \eqref{protodispersion} with \eqref{WEnp} using the Lemma to get $\cos\omega_\pm(k)=\frac12\tau_m(Wz(k))$. Thus in the expressions of the Lemma we have to make the replacements $a\mapsto \exp(ik)a$, $d\mapsto\exp(-ik)\overline{a}$ and $\det C\mapsto1$. Setting $a=\abs a\exp(ik_0)$, we get
\begin{align}\label{dispersEpaps}
  &\cos\omega_\pm(k)=\\
  &=\!\left\lbrace\!\! \begin{array}{cl}\abs a^{m}\cos(m (k+k_0))& m\mbox{\ odd}\\
                               -\abs a^{m}\cos(m (k+k_0))+(-1)^{{m}/2+1}(1-\abs a^m) &  m\mbox{ even}
                              \end{array}\right. \nonumber
\end{align}

\minisubsection{Proof of the revival theorem}
Intuitively, we will use that, according to the dispersion relations \eqref{dispersEpaps}, all $\omega_\pm(k)$ must be close to $\pm i$ in the odd case and close to $(-1)^(m/2)$ in the even case. To get the precise norm expressions, consider a normal operator $A$ (i.e., $[A^\dagger,A]=0$), which commutes with translations, so its action on momentum space wave functions is $(A\psi)(k)=A(k)\psi(k)$.
Then $\norm A=\max_k\norm{A(k)}=\max_{k,\pm}\abs{\alpha_\pm(k)}$, where $\alpha_\pm(k)$ are the eigenvalues of $A(k)$. We can apply this to the operators in the revival theorem. In the odd case,
\begin{align}
\opNorm{\,\We^{2\dr}+\idty\,}&=\max_{k,\pm}\left\vert e^{2i\omega_\pm(k)}+1 \right\vert     \nonumber\\
  &=\max_{k,\pm} \left\vert e^{i\omega_\pm(k)}+e^{-i\omega_\pm(k)}\right\vert\nonumber\\
  &=\max_{k,\pm} 2\cos{\omega_\pm(k)} \nonumber\\
  &=2\abs a^m
\end{align}
where at the last line we used that the maximum over $k$ is attained at $k=-k_0$. Similarly, in the even case we use that $\norm{A}^2=\norm{A^\dagger A}$ and get
\begin{align}
\opNorm{\,\We^{\dr}+(-1)^{\dr/2}\idty\,}^2
  \mskip-80mu&\strut\mskip80mu
    =\max_{k,\pm}\left\vert e^{i\omega_\pm(k)}+(-1)^{m/2}\right\vert^2      \nonumber\\
  &=\max_{k,\pm} \left(2+2(-1)^{m/2}\cos\omega_\pm(k)\right)\nonumber\\
  &=\Bigl(2(1-(1-\abs a^m)) +\Bigr.     \nonumber\\
  &\strut\quad \Bigl.+2\abs a^m \max_k (-1)^{m/2}\cos(m(k+k_0))\Bigr)\nonumber\\
  &=4\abs a^m
\end{align}

This concludes the proof of the revival theorem.
\fi

\minisection{Acknowledgements}
This project was supported by the DFG (Forschergruppe 635),the ERC grant DQSIM, and NRW Nachwuchsforschergruppe ``Quantenkontrolle auf der Nanoskala''. AA and MG also acknowledge support from the Alexander von Humboldt Foundation and the BCGS, respectively.

\pagebreak
\bibliography{ewalks}

\end{document}